\documentstyle[12pt]{article}
\newcommand{\bq}{\begin{equation}}
\newcommand{\eq}{\end{equation}}
\def\fr{\frac}
\def\be{\begin{equation}}
\def\ee{\end{equation}}
\begin{document}
%\voffset=-1.0truecm
%\hoffset=-1.0truecm

{\raggedleft{\sf${ ASITP}$-94-38\\}}
%{\raggedleft{\sf${ MPIM  }$-94\\}}

%{\raggedleft {\makebox[2.7cm][1]{\large\it~~~~ Draft}}\\}
\vskip 2mm
{\raggedleft {\makebox[2.7cm][1]{\it November 1994}}\\}
\vskip 10mm

\centerline{\Large\bf  A Note On Intrinsic  Regularization  Method}

\vskip 15mm
\begin{center}

{\large Han-Ying Guo{\footnote{Email: hyguo@itp.ac.cn}}}\\
\vskip 3mm
Max-Planck-Institut f\"ur Mathematik, D-53225 Bonn, Germany;\\
\vskip 2mm
and\\
\vskip 2mm
Institute of Theoretical Physics, Academia Sinica,
P.O. Box 2735, Beijing 100080, China.
\footnote {Permanent address.}\\

\vskip 6mm
{\large Yu Cai\footnote{Email: caiyu@itp.ac.cn} ~and~ Hong-Bo
Teng\footnote{Email: tenghb@itp.ac.cn} }\\
\vskip 2mm
Institute of Theoretical Physics, Academia Sinica,
P.O. Box 2735, Beijing 100080, China.\\
\end{center}
\vspace{8ex}
\vspace{8ex}

\centerline{\it ABSTRACT}
\vskip 1mm
\begin{center}
\begin{minipage}{142mm}
{\it There exist certain intrinsic relations between the ultraviolet divergent
graphs and the convergent ones at the same loop order in  renormalizable
quantum field theories. Whereupon we may establish a new method, the intrinsic
regularization method, to regulate
 those divergent graphs.  In this note, we present a proposal, the inserter
proposal, to the method.
 The  $\phi^4$ theory and QED at the
one loop order are dealt with in some detail. Inserters in the standard model
are
 given.  Some applications to SUSY-models are also made at the one loop order.}

\end{minipage}
\end{center}
\vskip 2pt
%~~~PACS:02.20.+b
\newpage

%\bigskip
%\bigskip
{\large\it  Introduction}
%\medskip

As is well known, various regularization  schemes
have been developed in the quantum field theory. However, the topic is
still one of the important and fundamental issues under investigation.
One of the most challenging problems is perhaps how to preserve  all
 symmetries and topological properties manifestly and consistently.

 It has been found that there exist certain intrinsic relations between
the  ultraviolet  divergent
graphs and the convergent ones at the same loop order in  renormalizable
QFT [1-5]. Whereupon we should be able to establish a new method, the intrinsic
regularization method,  for regularization
of those divergent graphs.
 In this note, we present a proposal to this
 method, the inserter proposal. We deal with the $\phi^4$  theory and QED
at the one loop order. We also make some applications to SUSY-models
 at the one loop order by means of  the SUSY version of the inserter proposal
 and explain how to apply it to other cases. This proposal may shed light on
that challenge.

The key point of the new method is, in fact,
based upon the following simple observation:
For a given  ultraviolet   divergent function at certain loop order in a
 renormalizable QFT, there always exists
a set of convergent functions at the same loop order such that their Feynman
graphs share the
same loop skeleton  and the main difference
is that the convergent ones have additional vertices of certain kind and the
original
one is the case without these vertices. This is, in fact, a certain intrinsic
relation between the original  ultraviolet  divergent graph and the convergent
ones in the
QFT. It is this relation that indicates it is possible to introduce the
regulated function for the divergent function with the help of those convergent
 ones so that the potentially divergent
integral of the graph can be rendered  finite while for the limiting case of
the
number of the additional vertices $q\to 0$
 the divergence again becomes manifest in pole(s) of $q$.

It is very simple why there always exists such kind of intrinsic relations in
renormalizable QFTs. Let us consider some Feynman graphs at the
$L$ loop order with  $I$ internal lines of any kind and  $V$ vertices of any
 kind. The topological formula
$$ L-I+V = 1$$
shows that for fixed $L$ $I$ increases the same as $V$ does so that the
superficial degree of divergence decreases. Therefore, for a given divergent
Feynman graph at certain loop order, the topological formula insures that one
can always reach a set of convergent graphs in a suitable  perturbation
 expansion series in the order of some coupling constant
, which always appears with a vertex of certain kind, as long as the original
divergent graph is included in the series. In fact, this topological formula is
a cornerstone of the intrinsic regularization method.  In general, however,
the procedure may be very involved. The aim of the inserter proposal to be
presented in this note is just to simplify the procedure.
 In fact, by means of
 this proposal
the fore-mentioned limiting procedure allows us to unambiguously calculate the
various
Feynman graphs.

To be concrete, let us consider a 1PI graph with $I$ internal lines at one loop
order in
the $\phi^4$ theory. Its superficial degree of divergences in the momentum
space is
$$
\delta=4-2I.
$$
When $I=1$ or $2$, the graph is divergent. Obviously, there exists such kind of
graphs that they have additional $q$ four-$\phi$-vertices in the internal
lines. Then the number of
 internal lines in these graphs is $I+q$  so that the
divergent degree of the new 1PI graphs become
%%\marginpar{\em changed}
$$
\delta'=4-2(I+q).
$$
If $q$ is large enough, the new ones are convergent and the original divergent
one is the case of $q=0$.  Thus,  a
certain intrinsic relation has been reached between the  original divergent 1PI
graph and the
new convergent ones at the same loop order.

In the inserter proposal,  we take all external
lines in the additional vertices with  zero momenta and call such a vertex
an {\it inserter}. Thus those convergent graphs can simply be regarded as the
ones given by suitably inserting $q$-inserters in all internal lines in the
given divergent graph and
 the powers of the propagators are simply raised so as to those integrals for
 the new  graphs become convergent.
It is clear that these new graphs share the same loop skeleton with the
original divergent one and the main differences between those graphs are the
number of inserters as well as the dimension in mass and the order in the
 coupling constant due to the insertions of the inserter. Thus
it is possible to regulate the
divergent graph based upon the intrinsic relations of this type as long as
 we may get rid of those differences and deal with those convergent
 functions on an equal footing. In fact, to this end we introduce a
  well-defined convergent 1PI function, the
 regulated function, by taking the
arithmetical average of those convergent 1PI functions and changing their
 dimension in mass, their order in $\lambda$ etc. to the ones in
the original divergent function.
 Thus this new function renders the divergent
integral in the original function finite. Evaluating it and continuing $q$
analytically from the integer to the complex number, the  divergent function
of the original 1PI
 graph is recovered as
the $q \to 0$
limiting case of such a regulated function.

It is not hard to see that
 in any given QFT as long as a suitable kind of inserters are constructed
with the help of the Feynman rules of the theory and some intrinsic relations
between the
divergent functions and convergent ones at the same loop order are found by
 inserting  the inserters, this  inserter proposal should work in principle. Of
course, special
attention should be devoted to each specified case.
In QED, for example, since the electron-photon vertex carries a $\gamma$-matrix
and is a Lorentz vector, therefore,
  how to construct a suitable inserter is the first problem to be solved
in addition to the above procedure in the $\phi^4$ theory. Otherwise, simply
inserting  the vertex would increase the rank of the functions as  Lorentz
tensors and the problem could become quite complicated. We will be back
to this point at the end of this note.
In order to avoid this complication, we borrow an inserter of the Yukawa type
for the massive fermions from the standard model
( see the appendix )
and employ
it for the purpose in QED. By inserting  this inserter to the
internal fermion lines in the graph of a given 1PI $n$-point divergent
 function, a set of new convergent functions can be obtained if the number of
 inserters, $q$, is large enough. Then we introduce a new convergent
 function, the regulated function, it not only should have the same dimension
in mass, and
the same order in the
coupling constant and so on as
 the original divergent function but also should preserve the gauge invariance.
 Thus the potential infinity in
the original 1PI $n$-point
function may be recovered as the $q\to 0$ limiting case of that function.

In what follows, we concentrate on  how to
regulate the divergent graphs at one loop order in
the $\phi^4$ theory and QED.
We present the main steps and the results
of the inserter regularization procedure for each of them.
 We also describe briefly  the SUSY version of the inserter proposal and make
some applications to SUSY-models
 at the one loop order.
 Finally, we end
with some discussion and remarks. In the appendix, different kinds of inserters
in the standard
model are given.

%\end{document}
\bigskip\bigskip
{\large\it   Intrinsic Regularization In $\phi^4$ Theory}
\medskip

 The action of the $\phi^4$ theory is
\begin{equation}
S[\phi]=\int d^4 x~(\fr{1}{2}\partial_{\mu}\phi\partial^{\mu}\phi -\fr{1}{2}
m^2 \phi^2 -\fr{\lambda}{4!}\phi^4)
\end{equation}
and its Feynman rules are well known.

 The main steps of the inserter proposal  for the $\phi^4$ theory
may more concretely be stated as
follows.
First, we should construct the inserter in the $\phi^4$ theory. As mentioned
above, it is a four-$\phi$-vertex with two zero momentum external lines whose
Feynman rule is the same as the vertex
\be
I^{\phi}(p)=-i\lambda.
\ee
For a given  1PI $n$-point
divergent function at the one loop order $\Gamma^{(n)}(p_1,~\cdots,~p_n)$,
 we consider all $n+2q$-point functions
$\Gamma^{(n+2q)}(p_1,~\cdots,~p_{n}; ~q)$ which are  the
amplitudes
of the graphs corresponding to all possible $q$ insertions of the inserter on
the internal lines
 of the given $n$-point graph.  If
$q$ is large enough, $\Gamma^{(n+2q)}(p_1,~\cdots,~p_{n}; ~q)$ become
convergent. And for $q=0$ it is the case of original
$n$-point function. This is a  relation between the given divergent
function and those convergent ones.  With the help of this relation, we
introduce a new function by taking the arithmetical average of these convergent
functions, i.e. the summation of these
 functions divided by ${N_q}$, the total number of such inserted
functions, and let it  have the same dimension in mass and the same order in
$\lambda$ as the original 1PI $n$-point function:
%%\marginpar{\em the phase factor is modified in this eq}
\be
\Gamma^{(n)}(p_1,~\cdots,~p_n; ~q; ~\mu)
=(-i\mu^2)^q(-i\lambda)^{-q}\fr 1 {N_q} \sum
\Gamma^{(n+2q)}(p_1,~\cdots~p_n;~q)
\ee
where $\mu$ is an arbitrary reference mass parameter. Note that this function
 is no longer an $n+2q$-point function rather a regulated $n$-point function
since it is at
 the same order in the coupling constant $\lambda$ as the original function.
Now we
 evaluate it and analytically continue $q$ from the integer to the
complex number.
Then the original potentially divergent 1PI $n$-point function is recovered by
\be
\Gamma^{(n)}(p_1,~\cdots,~p_n;\mu)=\lim_{q\to 0}
\Gamma^{(n)}(p_1,~\cdots,~p_n;~q;~\mu),
\ee
and the original infinity arises manifestly as pole in $q$.
Obviously, this procedure should in principle work for the cases at higher loop
orders.

 At the one loop order there are only two divergent graphs in the $\phi^4$
theory, the tadpole
$(t)$ and the fish $(f)$.
In the momentum space, the amplitude of $(t)$ and $(f)$ are
\begin{equation}
\begin{array}{ll}
(t)=\fr{1}{2}\int \fr{d^4 l}{(2\pi)^4} ~~\fr{\lambda}{l^2-m^2},\\[4mm]
(f)=\fr{1}{2}\int \fr{d^4 l}{(2\pi)^4}~~
\fr{\lambda^2}{(l^2-m^2)((p_1+p_2+l)^2-m^2)} +(p_2 \to p_3) +(p_2 \to p_4).
\end{array}
\end{equation}
They are quadratically
 and logarithmically divergent respectively and needed to be
regulated.

In order to regulate the tadpole, we  attach $q$ inserters to
the internal line  of the graph. Then $(t)$ becomes a $2+2q$-point
function $(t_q)$. For $q$ large enough, $(t_q)$ is
convergent. We now introduce a new function $(t_q')$ which has the same
 dimension in mass and the same order in $\lambda$ with $(t)$ and
when $q=0$,
$(t_q')\vert_{q=0}=(t)$.

The amplitude of $(t_q')$ can be expressed as {\footnote{In this note,
the order of the inserted inserters in each inserted graph is always fixed so
that the relevant combinatory factor is  simply fixed to be one as well.}}
%%\marginpar{\em the phase factor is modified}
\begin{equation}
(t_q')=(-i\mu^2)^q(-i\lambda)^{-q}(t_q)=\fr{1}{2}\mu^{2q}\int \fr{d^4
l}{(2\pi)^4}
\fr{\lambda}{(l^2-m^2)^{q+1}}.
\end{equation}
It can be easily integrated and expressed
in terms of the $gamma$ functions of $q$:
\begin{equation}
(t_q')=\mu^{2q}\fr{i}{2}\fr \lambda {(4\pi)^2}\fr{\Gamma(q-1)}
{\Gamma(q+1)(-m^2)^{q-1}}.
\end{equation}
Now we analytically continue
  $q$ from the integer to the complex number. The original
 tadpole function $(t)$ is then recovered as  the $q\to 0$ limiting case
of $(t_q')$:
\begin{equation}
(t)=\lim_{q\to 0} (t_q')=\fr{i}{2}\fr \lambda {(4\pi)^2}m^2[\fr{1}{q}
+1+\ln (-\fr{\mu^2}{m^2})+o(q)].
\end{equation}
%Obviously, the regularization of the tadpole function is completed.

In order to regulate the fish, we attach to its internal lines
$q$ inserters and
it turns to a set of the graphs $(f_{q,i})$ with $i$
inserters inserted on one internal line while $q-i$ on the other.
For $q$ large enough, all $(f_{q,i})$ are convergent.
Let us introduce their arithmetical average
$(f_q)=\fr 1 {N_q}\sum_{i=0}^q(f_{q,i}), ~{N_q}=q+1$ and a new
function $(f_q')$ which has
the same dimension in mass and the same order in $\lambda$ with that  of
$(f)$ and when $q=0$, $(f_q')\vert_{q=0}=(f)$.

The amplitude of $(f_q')$ can be expressed as
%\marginpar{\em the phase factor is modified}
\begin{equation}\begin{array}{cl}
(f_q')&=(-i\mu^2)^q(-i\lambda)^{-q}(f_q)\\[4mm]
&=\fr{\mu^{2q}}{2(q+1)}\sum_{i=0}^{q}\int \fr{d^4 l}{(2\pi)^4}
%% FOLLOWING LINE CANNOT BE BROKEN BEFORE 80 CHAR
\fr{\lambda^{2}}{(l^2-m^2)^{i+1}((p_1+p_2+l)^2-m^2)^{q-i+1}}\\[4mm]
&~~~+(p_2 \to p_3) + (p_2\to p_4).\end{array}\end{equation}
And we can  get
\begin{equation}
\begin{array}{ll}
(f_q')
=\fr{i}{2(4\pi)^2}\lambda^{2}\mu^{2q}\fr{1}{q(q+1)}
\int_0^1d\alpha \fr{1}{[\alpha(1-\alpha)(p_1+p_2)^2-m^2]^q}
+(p_2 \to p_3) + (p_2\to p_4).\end{array}
\end{equation}
We now  analytically continue $q$ from the
integer to the complex number. Then the fish function
$(f)$ is reached by  the $q\to 0$
limiting case of $(f_q')$,
%\marginpar{\em modified}
\begin{equation}
\begin{array}{cl}
(f)&=\lim_{q\to 0}(f_q')\\
&=i\fr {{\lambda}^2}
{(4\pi)^2}[\fr{3}{2q}+\fr{3}{2}+\fr{3}{2}\ln(-\fr{\mu^2}{m^2})+
A(p_1,\cdots, p_4)+o(q)],
\end{array}
\end{equation}
where
$$A(p_1,\cdots, p_4)=
-\fr{1}{2}\sqrt{1-\fr{4m^2}{(p_1+p_2)^2}}
\ln \fr{\sqrt{1-\fr{4m^2}{(p_1+p_2)^2}}+1}{\sqrt{1-\fr{4m^2}{(p_1+p_2)^2}}-1}+
(p_2 \to p_3) + (p_2\to p_4).$$

Thus we complete the regularization of the $\phi^4$ theory at the one loop
order by means of the  inserter proposal.
%\end{document}

\bigskip\bigskip
{\large\it  Intrinsic Regularization In QED}
\medskip

The action of the QED is
\be
S[{\bf A}, \psi]=\int d^4 x \{
-\fr{1}{4}F_{\mu\nu}F^{\mu\nu}
-\fr{1}{2 \xi}(\partial \cdot {\bf A})^2+
\bar{\psi}(i/\!\!\!{\partial}-m)\psi-e \bar{\psi} /\!\!\!\!{A}{\psi}\}.
\ee
and the Feynman rules are well known.

As was mentioned above, we first employ an {\it inserter} borrowed from the
standard model. It is an $ff\phi$-vertex of the Yukawa type with a zero
 momentum Higgs external line.  The Feynman rule of such an inserter is
\be
I^{\{f\}}(p)=-i\lambda_f,
\ee
where $\lambda_f$ takes value $\fr g 2  {m_f} /{M_W}$ in the standard model,
but
here its value is irrelevant for our purpose. Then for a given  divergent 1PI
amplitude
$\Gamma^{(n_f,n_g)}_{~ *}
 (p_1,\cdots, p_{n_f};k_1,\cdots, k_{n_g})$ of rank $*$ Lorentz tensor at the
one loop order with $n_f$
external fermion lines and $n_g$ external photon lines, we consider a set
of 1PI amplitudes
$
\Gamma^{(n_f,n_g, 2q)}_{~ *} (p_1,\cdots, p_{n_f};k_1,\cdots, k_{n_g}; q)
$
which correspond to the graphs with all possible $2q$ insertions of
the inserter in the internal fermion lines in the original graph.
Because
each insertion  decreases the divergent degree by $1$, the divergent
degree  becomes:
%\marginpar{\em changed}
$$
\delta=4-(I_f+2q)-2I_g.
$$
If $q$ is large enough, $\Gamma^{(n_f,n_g, 2q)}_{~ *} (p_1,\cdots,
p_{n_f};k_1,\cdots, k_{n_g}; q)$ are
convergent  and the original divergent function is the
 case of $q=0$. Thus we reach a relation
between the given divergent 1PI function and a set of convergent 1PI functions
at the one loop order. In fact, the function of inserting such an inserter to
an internal fermion line is simply to raise the power of the
propagator of the line and to decrease the degree of divergence of given
graph.

In order to regulate the given divergent function with the help of this
 relation, we need to deal with those convergent functions on an
 equal footing and pay attention to their differences due to the insertions.
To this end, we introduce a new
function:
%\marginpar{\em the phase factor is modified}
\be
\begin{array}{ll}
\Gamma^{(n_f,n_g)}_{~ *} (p_1,\cdots, p_{n_f};k_1,\cdots, k_{n_g}; ~q;~\mu)
{}~~~~~~\\[4mm]
{}~~~~~~= (-i\mu)^{2q} (-i\lambda_f)^{-2q} \fr 1 {N_q} \sum \Gamma^{(n_f,n_g,
2q)}_{~ *}
(p_1,\cdots, p_{n_f};k_1,\cdots, k_{n_g}; ~q)
%\eta^{\mu_1 \mu_2} \cdots\eta^{\mu_{2q-1} \mu_{2q}},
\end{array}
\ee
where $\mu$ is an arbitrary reference mass parameter as in the last section,
the factor $(-i\lambda_f)^{-2q}$ introduced here is to cancel the one coming
from
$2q$-inserters and the summation is taken over
the entire set of such ${N_q}$ inserted functions.
It is clear that this function is the arithmetical average of  those
convergent functions and has  the same dimension in mass, the same order in $e$
 with the original divergent 1PI function.
 Now we
 evaluate it and continue analytically $q$ from the integer to the complex
number.
Finally, the  original 1PI function should be recovered as its $q\to 0$
limiting case:
\be
\Gamma^{(n_f,n_g)}_{~ *} (p_1,\cdots, p_{n_f};k_1,\cdots, k_{n_g})
=\lim_{q\to 0}
\Gamma^{(n_f,n_g)}_{~ *} (p_1,\cdots, p_{n_f};k_1,\cdots, k_{n_g};q;\mu),
\ee
and the original infinity would appear as pole in $q$. Similarly, this
procedure should work for the cases at the higher loop orders in principle.
%\end{document}

The divergent 1PI graphs at the one loop order in QED are those contribute
to the
vacuum polarization $\Pi_{\mu \nu}(k)$, the electron self-energy $\Sigma(p)$,
the vertex function $\Lambda_{\mu} (p',p)$ and the photon-photon scattering
function
$\Gamma_{ \mu \nu \rho \sigma } (p_1, \cdots, p_4)$. Their
integral expressions
 in the momentum space are as follows ( For simplicity, we take the
Feynman gauge $\xi =1$. ):
%\marginpar{\em modified}
$$\Pi_{\mu \nu}(k)=-e^2\int \fr{d^4
p}{(2\pi)^4}Tr[\gamma_{\mu}\fr{1}{\not{p}-\not{k}-m}
\gamma_{\nu}\fr{1}{ \not {p} -m}],$$
$$\Sigma(p)_{\beta\alpha}=-e^2\int \fr{d^4 k}{(2\pi)^4}
(\gamma^{\mu}\fr{1}{ \not{k}-m}\gamma_{\mu})_{\beta\alpha}
\fr{1}{(p-k)^2},
$$
\be
%{}~~~~~~~~~~~
\Lambda_{\mu} (p',p)_{\beta\alpha} =-e^3\int \fr{d^4
l}{(2\pi)^4}(\gamma^{\rho}\fr{1}{\not{l}-\not{k}-m}
\gamma_{\mu}\fr{1}{\not{l} -m}
\gamma_{\rho})_{\beta\alpha} \fr{1}{(p-l)^2},
\ee
$$\begin{array}{cl}
\Gamma_{ \mu \nu \rho \sigma } (p_1, \cdots, p_4)
&=
-e^4\int \fr{d^4 k}{(2\pi)^4}
Tr\{\gamma_{\mu}\fr{1}{\not{k}-m}
\gamma_{\nu}\fr{1}{\not{k}+\not{p_2} -m}
\gamma_{\rho} \fr{1}{\not{k} +\not{p_2} +\not{p_3}-m}
\gamma_{\sigma} \fr {1} {\not{k}-\not{p_1}-m}\}\\[4mm]
&~~+(\mu \leftrightarrow \nu, p_1 \leftrightarrow p_2)+
(\mu \leftrightarrow\rho, p_1 \leftrightarrow p_3)\\[4mm]
&~~+(\mu \leftrightarrow \sigma, p_1 \leftrightarrow p_4)
+(\nu \leftrightarrow \rho, p_2 \leftrightarrow p_3)\\[4mm]
&~~+(\rho \leftrightarrow
\sigma, p_3 \leftrightarrow p_4)
\end{array}
$$
which are superficially quadratically, linearly and
logarithmically ultraviolet divergent respectively.
Let us now render them finite by means of the inserter procedure.

To regulate the divergent vacuum polarization function $\Pi_{\mu \nu}(k)$,
we attach to one
internal fermion line with $i$ inserters  and to
the other  with $2q-i$ ones. Then we get
a set of $2+2q$-point functions $\Pi^{\{q,i\}}_{\mu \nu}(k; q)$. If $q$ is
large enough, all these
$2+2q$-point functions are convergent. Then we introduce a new function
%\marginpar{\em phase is modified}
\be
\Pi_{\mu \nu}(k;q;\mu)=(-i\mu)^{2q} (-i\lambda_f)^{-2q} \fr 1 {N_q}
\sum_{i=0}^{2q}
\Pi^{\{q,i\}}_{\mu \nu }(k; q),
%\eta^{\mu_1 \mu_2} \cdots\eta^{\mu_{2q-1} \mu_{2q}},
\ee
which has the same dimension in mass, the same order in $e$ with the original
function $\Pi_{\mu
\nu}(k)$. It is not hard to prove that this  function can be  expressed as
$$
\Pi_{\mu \nu}(k;q;\mu)=-\mu^{2q}e^{2}\fr 1 {N_q} \sum_{i=0}^{2q}\int \fr{d^4
p}{(2\pi)^4}
Tr[\gamma_{\mu}\Bigl(\fr{1}{\not{p}-\not{k}-m}\Bigr)^{i+1}
\gamma_{\nu}\Bigl(\fr{1}{ \not {p} -m}\Bigr)^{2q-i+1}]
$$
and satisfies the gauge invariant condition:
\be
k^{\mu}\Pi_{\mu \nu}(k;q;\mu)=0.
\ee
Continuing $q$ to the complex number, thus the original amplitude
$\Pi_{\mu\nu}$ is recovered as
\be
\Pi_{\mu\nu}(k)=\lim_{q\to 0}\Pi_{\mu\nu}(k;q;\mu).
\ee
If we denote
\be
\Pi_{\mu\nu}(k^2)\equiv
(k_{\mu}k_{\nu}-k^2g_{\mu\nu})\Pi(k^2),~~~\Pi(k^2)=\Pi(0)+\Pi^f(k^2),
\ee
by some calculation, we get
\be\begin{array}{ll}
\Pi(0)
=e^2\fr{4i}{(4\pi)^2}[\fr{1}{3q}
+C+\frac{1}{3}\ln(-\fr{\mu^2}{m^2}) +
o(q)],\\[4mm]
\Pi^f(k^2)=-\fr{ie^2}{2\pi^2}
\int_0^1 d\alpha \alpha(1-\alpha)\ln[1-\fr{\alpha(1-\alpha)k^2}{m^2}].
\end{array}\ee
where $C$ is some constant. The finite part $\Pi^f(k^2)$ is the same as that
derived in other regularization procedures.

To regulate the electron self-energy function $\Sigma(p)$, we attach to the
internal fermion line with $2q$ inserters and the
graph $\Sigma(p)$ is turned to a $2+2q$-point convergent function
 $\Sigma(p; q)$ if $q$ is large enough. Then we introduce a
new function:
%\marginpar{\em phase modified}
\be
\Sigma(p; q;\mu)=(-i\mu)^{2q} (-i\lambda_f)^{-2q}\Sigma (p;q),
%\eta^{\mu_1\mu_2} \cdots \eta^{\mu_{2q-1}\mu_{2q}},
\ee
which can be expressed as
%\marginpar{\em modified}
$$
\Sigma(p; q;\mu)_{\beta\alpha}=-\mu^{2q}e^2
\int \fr{d^4 k}{(2\pi)^4}
\Bigl(\gamma^{\mu}\fr{(\not{k}+m)^{1+2q}}{(k^2-m^2)^{1+2q}}
\gamma_{\mu}\Bigr)_{\beta\alpha}\fr{1}{(p-k)^2}.
$$
Continuing $q$ to the complex number, the original function  $\Sigma(p)$ is
 reached by
\be
\Sigma(p)_{\beta\alpha}=\lim_{q\to 0}\Sigma(p; q;\mu)_{\beta\alpha}.
\ee
Denoting
\be
\Sigma(p)=mA(p^2)+iB(p^2)\not{p},
\ee
we may finally get
$$
\begin{array}{cl}
A(p^2)=&-\fr{ie^2}{4\pi^2}\{\fr{1}{q}+3-\ln\fr{p^2-m^2}{\mu^2}
+\fr{m^2}{p^2}\ln(1-\fr{p^2}{m^2})+A^f(p^2)\},\\[4mm]
B(p^2)=&\fr{e^2}{(4\pi)^2}\{\fr{1}{q}+\fr{1}{2}+\fr{m^2}{p^2}
-\ln\fr{p^2-m^2}{\mu^2}
-(\fr{m^2}{p^2})^2\ln(1-\fr{p^2}{m^2})\}+B^f(p^2),
\end{array}
$$
both $A^f(p^2)$ and $B^f(p^2)$ are finite function.
%\end{document}

Similarly,  to regulate the  vertex function,
 $\Lambda_\mu (p', p)$, we attach to the internal fermion lines with $2q$
inserters to  get a set of $(3+2q)$-point functions
$\Lambda^{\{q,i\}}_\mu(p', p; q)$  with $i$ inserters on one
internal fermion line and $q-i$ inserters on the other.
Then we introduce a new function
%\marginpar{\em phase modified}
\be
\Lambda_\mu(p', p; q;\mu)=(-i\mu)^{2q} (-i\lambda_f)^{-2q}\fr 1 {N_q}
\sum_{i=0}^{2q}
\Lambda^{\{q, i\}}_{\mu}(p',p; q),
%\eta^{\mu_1\mu_2 }\cdots\eta^{\mu_{2q-1}\mu_{2q}},
\ee
which is convergent if $q$ is large enough and has the same dimension in mass,
the
same order in $e$ with the original vertex function $\Lambda_\mu(p', p)$. It
can be expressed as
$$
\Lambda_\mu(p', p; q;\mu)=-\fr {\mu^{2q}e^{3}} {2q+1} \sum_{i=0}^{2q}
\int \fr{d^4 l}{(2\pi)^4}
\Bigl(\gamma^{\rho}\fr{(\not{l}-\not{k}+m)^{1+i}}
{[(l-k)^2-m^2]^{i+1}}\gamma_{\mu}
\fr{(\not{l}+m)^{1+2q-i}}{(l^2-m^2)^{1+2q-i}}\gamma_{\rho}
\Bigr)
%_{\beta\alpha}
\fr{1}{(p-l)^2}.
$$
Continuing $q$ to the complex number,  the original vertex function is then
recovered
as:
\be
\Lambda_\mu(p', p)=\lim_{q\to 0}\Lambda_\mu(p', p; q;\mu).
\ee
Finally, we find that
\be\begin{array}{ll}
\Lambda_\mu(p', p)
=\fr{-ie^3\mu^{2q}}{(4\pi)^2}\gamma_{\mu}~~~~~~~~~\\[4mm]
{}~~[\fr{1}{q}-\fr{3}{2}-
\int_0^1 d\alpha\int_0^{1-\alpha}d\beta
\ln \{\beta(1-\beta)k^2+\alpha(1-
\alpha)p^2-2\alpha\beta k\cdot p-(1-\alpha)m^2\}]\\[4mm]
{}~~-\fr{ie^3}{2(4\pi)^2}\int_0^1 d\alpha\int_0^{1-\alpha}d\beta \fr
{\gamma^{\rho}[(\beta-1)\not{k}+\alpha \not{p}+m]\gamma_{\mu}
[\beta\not{k}+\alpha\not{p}+m]\gamma_{\rho}}{\beta(1-\beta)k^2+\alpha(1-
\alpha)p^2-2\alpha\beta\cdot p-(1-\alpha)m^2}+o(q).
\end{array}
\ee
The observable part $\Lambda_{\mu}^f$  of the vertex function is
defined by
$$
\Lambda_{\mu}=K\gamma_{\mu}+\Lambda_{\mu}^f,
$$
where $K$ contains the pole in $q$ when $q\to 0$ and $\Lambda_{\mu}^f$ is
finite from which the anomalous magnetic moment of the
electron can be derived. The result is the same as in other approaches.
%\end{document}

Similar procedure may also be applied to
the photon-photon scattering
$\Gamma_{\mu \nu \rho \sigma}(p_1 \cdots p_4)$.
We check its gauge invariance
by the inserter proposal.
Attaching $2q$ inserters to internal fermion lines in all possible ways
we get a
set of convergent functions if $q$ is large enough. Then we introduce a new
function
\be
%\begin{array}{ll}
\Gamma_{\mu \nu \rho \sigma}(p_1 \cdots p_4;q;\mu)
=(-i\mu)^{2q} (-i\lambda_f)^{-2q}
%~~~~~~~{}\\[4mm]
%{}~~~~~~
\fr 1 {N_q}\sum_{i=0}^{2q}\sum_{j=0}^{2q-i}
\sum_{l=0}^{2q-i-j} \Gamma^{\{q,i,j,l\}}_{\mu \nu \rho \sigma }
(p_1 \cdots p_4),
%\eta^{\mu_1\mu_2} \cdots \eta^{\mu_{2q-1}\mu_{2q}},
%\end{array}
\ee
which has required properties with respect to the  original
 function.
 It can be proved that this function satisfies the gauge invariant condition.
Continuing
 $q$ to the complex number, the
 original  function is then recovered by
\be
\Gamma_{\mu \nu \rho \sigma}(p_1 \cdots p_4)
=\lim_{q\to 0}\Gamma_{\mu \nu \rho \sigma}(p_1 \cdots p_4; q;\mu).
\ee
By some straightforward calculation, we may explicitly show that
\be
\Gamma_{\mu \nu \rho \sigma}(p_1 \cdots p_4)\vert_{p_1= \cdots =p_4=0}=0.
\ee
This also coincides with the gauge invariance.

Thus we complete the regularization of QED at the one loop order by means of
 the  inserter proposal.

\bigskip\bigskip
{\large\it Some Applications to SUSY-Models}
\medskip

We now apply the inserter proposal for the intrinsic regularization to some
SUSY-models at one-loop order. We will not present any detailed calculation
here.
The aim is to show that the SUSY version of the inserter proposal should work
and  preserve
supersymmetry manifestly and consistently by reexamining some well-known and
simple examples at
the one loop order.

Let us first consider an example in the massive Wess-Zumino model.
It is well
 known that at the one
loop level, the self-energy graph of antichiral-chiral superfield propagator
$\bar\phi \phi$
 is divergent. After some $D$-algebraic manipulation, it is left a divergent
 integral
\be
\int d^4\theta ~\phi(-p, \theta) \bar \phi(p, \theta)~A(p, m),
\ee
where
\be
A(p, m)
=\int \fr{d^4 k}{(2\pi)^4}
\fr{1}{(k^2+m^2)((p+k)^2+m^2)}.
\ee

To regulate this integral by means of the inserter proposal, we need first to
 construct an antichiral-chiral superfield inserter. For such an inserter, we
take a pair of vertices linked
by a $\bar\phi \phi$-internal line with a pair of chiral and antichiral
external
legs carrying zero momenta. Its Feynman rule can easily be written down.
Now we may utilize this inserter to insert $q$-times the internal lines in the
divergent graph. Then we get a set of convergent graphs with $i$-inserters on
one internal line and $q-i$ on the other. Similarly, after  some $D$-algebraic
manipulation, the corresponding convergent function $I^{\{q,i\}}(p)$ is
proportional to
\be
\int d^4\theta~\phi(-p, \theta)  \bar \phi(p, \theta)~A^{\{q,i\}}(p, m),
\ee
where
\be
A^{\{q,i\}}(p, m) =\int \fr{d^4 k}{(2\pi)^4}
\fr{1}{(k^2+m^2)^{2i+1}((p+k)^2+m^2)^{2q-2i+1}}.
\ee
It is very similar to the ones in the case of inserted fish functions, accept
 the sign of the mass due to the convention. Now we may almost repeat
the procedure in the $\phi^4$ theory to define the regulated function and so
on.
The original divergent function is recovered in the limiting case of $q \to 0$
and the divergence manifestly appears as a pole of $q$. It is so analogous to
the case of the fish in the $\phi^4$ theory that we do not need to repeat it
 here.

It is easy to see that the SUSY version of the inserter proposal may also
be applied to the massless
Wess-Zumino model  as well as other models.  Let us now consider a most general
$N=1$ supersymmetric  renormalizable model invariant under a gauge group $G$
contains chiral superfields $\phi^a$ in a representation $R$ of $G$ and $N=1$
Yang-Mills field contained in the general superfield $V$.

The one-loop correction to the $\phi^a \bar \phi_a$ propagator are given by two
divergent graphs. One is the same as in the Wess-Zumino model while the other
is the one with an internal antichiral-chiral line replaced by a $V$-line.
They lead to the expression:
\be
g^2 \int d^4\theta ~\phi^b(-p, \theta)
(S^a_b-C_2(R) \delta^a_b) \bar \phi_a(p, \theta)~A(p),
\ee
where $A(p)=A(p, m=0)$ and
$$ 2g^2 S^a_b=d^{bce}d_{ace},$$
$d^{bce}$ are the couplings of  the $\phi^3$-term. the cancellation condition
$S^a_b=C_2(R) \delta^a_b$ should hold much safer if the divergent integrals
 $A(p)$
in two graphs can be regulated in a way of preserving supersymmetry manifestly
and consistently. This can be done by means of the SUSY-version of the inserter
proposal.
To this end, in addition to the inserter for the chiral superfield constructed
above ( the internal representation indices should be paired here ),
we need an inserter for the $V$-internal line as well. In fact,
it may be constructed in such a way
that two $V\phi \phi$-vertices linked by an internal antichiral-chiral line
with two external
$\phi^a$, $\bar \phi_b $ legs carrying zero momenta and paired representation
indices. Then it is easy to see that by inserting these two inserters to the
internal $V$ line and the internal
  antichiral-chiral line respectively, we can always get the same regulated
functions for the both graphs. Therefore, the cancellation can be insured at
the regulated function level as well.

 The one-loop correction to the vector superfield propagator is given by three
 graphs with $V$-loop, $\phi$-loop and the ghost loop respectively. The SUSY
version of the inserter proposal also ensures the corresponding cancellation
condition holds at the regulated function level as long as we employ the ghost
inserter as a pair of the  ghost-V vertices linked by an internal ghost
antichiral-chiral  line with
two external $V$ legs carrying zero momenta in addition to the fore-mentioned
inserters for the internal $V$ line and the internal antichiral-chiral line.

The SUSY version of the inserter proposal may also be combined with the
background approach. For example, in the background field approach
the above one-loop contribution to the $V$ self-energy from a massive chiral
superfield
leads to one divergent integral only:
\be
\fr 1 4 C_2 ({R})~ tr
\int d^4\theta ~W^\alpha (p, \theta ) \Gamma_\alpha( -p, \theta)~A(p, m),
\ee
where $W^\alpha (p, \theta )$ is the superfield strength and $ \Gamma_\alpha(
-p, \theta)$ the background field connection. Again, we may utilize the
an antichiral-chiral superfield inserter
 in the background field to get a set
of convergent integrals and to regulate this divergent one in the way of
preserving supersymmetry manifestly and consistently.

It should be noticed that all construction for the superfield inserters and
regularization
for the divergent graphs are made with the help of the super-Feynman rules.
It is
 natural to expect that  the SUSY version of the inserter proposal does
preserve
supersymmetry manifestly and consistently not only for the one-loop cases but
also for the high-loop cases. We will explore this issue in detail
elsewhere.

\bigskip\bigskip
{\large\it Further Remarks}
\medskip

We have shown the main steps and results for  the regularization of the
divergent 1PI functions at the one loop order in both $\phi^4$ theory and QED
by means of the inserter proposal for the intrinsic regularization method.
 Some applications to SUSY-models are also made at the one loop order by means
of
 the SUSY version of the inserter proposal.
The results are satisfactory. It is naturally to expect that this proposal
 should  be available to the cases at higher loop orders in principle.

 The crucial point of
this approach, in fact, is very simple but fundamental. That is, the entire
procedure
 is intrinsic in the QFT. There is nothing
changed, the action, the Feynman rules, the spacetime dimensions etc.
are all the same as that in the given QFT. Although for QED, the inserter we
have employed is borrowed from the standard model, QED is in fact unified with
the weak interaction in the standard model. Therefore, it is still intrinsic
in the standard model. Consequently, in applying to other cases all
symmetries and
topological properties there should be preserved  in principle.
This is a very important property which should shed light on that challenging
problem. It is reasonable to expect that this proposal should be  able to apply
 consistently to those cases
where the symmetries and topological properties are sensitive to the
spacetime dimensions,
the number of
 fermionic degrees of freedom, such as chiral symmetry, anomalies,
SUSY theories etc.. As was shown in the last section, it is the case for some
SUSY-models at the one loop order.
Of course, for each case some special care should be taken. For the non-Abelian
gauge theories, like QCD and the standard model, for instance, special
attention
 should be
devoted  to the Lorentz indices and those indices of
 the internal gauge symmetries in constructing  the inserters. The Lorentz
 indices can be handled by contracting pairly by the spacetime metric.
Similarly,
 the internal gauge symmetry indices may also be dumbed
 by the Killing-Cartan metrics in the
 corresponding representations.
In the appendix, we  construct the inserters in the standard model. It
is straightforward to apply them at the one loop order. For
higher loop orders and other theories,
we will study them in detail elsewhere.

The renormalization of the QFT under consideration in this scheme should be the
 same as in  usual
approaches. Namely, we may subtract the divergent part of the $n$-point
functions
at each loop order by adding the relevant counterterms to the action.
 The renormalized $n$-point
functions are then evaluated from the renormalized action.
In the limiting case, we
get the finite results for all correlation functions.

In our proposal the inserters play an important role. The zero-momentum-line(s)
in the inserters do not of
course correspond to realistic particles. But, it  may have some physical
explanation. Namely, for each inserted internal line, the virtual particle
 always
emits
 and/or absorbs  via the
 inserters other far-infrared ``particles'' that carry zero momenta from
the vacuum. In other wards, the vacuum
is full of such far-infrared ``particles'' that they always have or pair
 together with the vacuum quantum numbers, i.e.  zero momenta,  singlet(s) in
all internal symmetries ( including gauge
symmetries ) and scalar(s) in the spacetime symmetries.
The ill-definess of
 those divergent
graphs can be handled by taking into account the role played by these
 far-infrared
``particles''.
This is just what has been done in the intrinsic inserter proposal.
An analogous explanation may also be made for the SUSY version of the inserter
proposal in terms of the superspace and superfields.

 We have not devoted any attention in this note to the infrared divergences at
all. It is in fact another most challenging problem to the regularization
schemes. In the course of
application to QED, the internal photon lines are the same as the original
ones. It is intriguing to see, however, as far as the vacuum picture
 is concerned,  certain kind of inserters should be constructed and some
intrinsic relation between
the divergent function and convergent ones may also be established in the
 infrared region. Then
the intrinsic inserter proposal may work in this region as well. We will
also
investigate this issue elsewhere.

On the other hand, however, as was fore-mentioned, although the
 inserter proposal works
 it is still a simplified procedure from the point of view of the intrinsic
regularization method. In fact, what have been taken into account is
 a proper set of all
convergent 1PI functions which share the same loop skeleton with the given
divergent 1PI functions. Not all of them.
It is obvious that the  simplification of the inserter proposal
certainly leads to a
 question  what role should be played by other convergent functions which can
 not be given by
 inserting  the inserters.
As a matter of fact, we may propose an alternative
 approach to the method.
For the $\phi^4$ theory and for some SUSY-models, for example, the ( massive )
Wess-Zumino model, it is the same as the inserter proposal or its SUSY-version.
 But for QED,
it is different: In spite of the complication mentioned before,
 we first simply attach $2q$ fermion-photon vertices with zero
momentum photon lines to the internal fermion line(s) in the graph of
given divergent 1PI $n$-point function with rank $r$ as a Lorentz tensor. By
doing so, if $q$ is large enough, we can  get a set of
 convergent 1PI $n+2q$-point functions
of rank $r+2q$  Lorentz tensors. And the original one is the case of $q=0$.
Thus we also reach an intrinsic relation between the original divergent
function and
those convergent ones. Although not only they are  at different order in the
coupling constant $e$ but
 also they have different ranks as Lorentz tensors, it is still possible
 to define a regulated function with the help of this relation. To this end,
we may
take all possible ways of contracting those additional Lorentz indices
in convergent functions  by the spacetime metric
to reduce
 the rank of Lorentz tensor to  the one in the original Lorentz tensor
function.
 Then we may employ the same procedure as that in the inserter proposal
to introduce the regulated function.
It is easy to see that the total number of the convergent
graphs in this approach could be different from and larger than that in the
inserter proposal. While the total number of the convergent functions after
contracting
the additional Lorentz indices is even much bigger than the one in the inserter
 proposal. Of course, this approach is much more complicated than the inserter
proposal and the calculation is also more tedious. But, as far as the
topological formula and those intrinsic relations are
concerned,   this alternative
approach
may also be available. Thus,
it is of course  interesting to see
whether there are some essential
differences between these two approaches.
We would leave this
classification problem for further investigation.

Finally, it should be mentioned that some idea of the inserter proposal was
first presented for the $\phi^4$ theory in [1] by ZHW and HYG as what is called
the intrinsic vertex
regularization. Later, the intrinsic loop regularization method has been
studied in [2-5]. Most results for the $\phi^4$ theory and QED at the one
loop order in [2,3,5] are very similar to what have been given in this note.
The mass
shifting in that approach, however, is not really intrinsic and do not
completely work
 for the theories with self-interacting massless particles, like QCD, the
standard model, SUSY-models etc.. The approach presented here should be able to
 get rid
 of all those problems.

\bigskip\bigskip
{\large\it Appendix:  Inserters In The Standard Model  }
\medskip

In order to apply the intrinsic inserter proposal to SM, different kinds of
 inserters are needed for inserting the internal lines of quarks, leptons,
gauge
bosons, Higgs and
 ghosts in the Feynman graphs with divergent amplitudes to be regulated.
To construct appropriate inserters we choose suitably a vertex or a pair of
vertices linked by an internal line and make merely use of the Feynman rules in
SM. In all inserters, the external leg or legs are all being managed in such a
way that they always carry
the vacuum quantum numbers, i.e. zero momentum, singlet in internal
and gauge symmetries,
and scalar in the spacetime symmetry. For
some specified internal line(s), different inserter may be employed for
different
 purpose.

1. The fermion-inserters:

There are two types of inserters. The Yukawa-inserters for massive fermions and
the one
for the neutrinos. For the Yukawa-inserters, we take them as corresponding
$ff\phi$-vertices
with zero-momentum Higgs lines. The Feynman rule is:
$$ I^{\{f\}}(p)=-i\fr g 2 \fr {m_f} {M_W}.$$
For each neutrino-inserter, we take a pair of $\nu_l\nu_l Z$-vertices linked by
an internal neutrino line such that two $\gamma$-matrices are contracted and
$Z$-external lines carry zero momenta. The Feynman rule is then:
$$ I^{\{\nu_l\}} (p)=\fr {g^2} {4 cos^2\theta_W} \fr i {\not{p}+i\epsilon}
(1-\gamma_5).$$

2. The gauge-boson-inserters:

For the gauge bosons such as gluons, $W^{\pm}$ and $Z$, there are some options.
We may take an inserter as a
4-gauge-boson vertex with two  zero-momentum lines whose indices are dumbed by
the spacetime metric and the Killing-Cartan metric of the gauge algebra
respectively. Their Feynman rules are easily be given. For example, for the
gluon-inserter:
$$I^{\{g\}ab}_{~~\mu\nu}(p)=-6ig_c^2C_2({\bf 8}) g_{\mu\nu} \delta^{ab},$$
where $C_2({\bf 8})$ is the second Casimir operator valued in the adjoint
representation
of $SU_c(3)$ algebra. On the other hand, we may also take a pair of
 3-gauge-boson vertices linked by an internal gauge-boson line with two
zero-momentum lines and dumbed pairly indices.

3. The Higgs-inserters:

Similar to the one in the $\phi^4$ theory, each Higgs-inserter may be taken
as a suitable 4-$\phi$-vertex with two zero momentum lines. Their Feynman rules
are easily be given as well. For example, the inserter for $\phi^i, i=1,2,$
is
$$ I^{\{\phi^i\}}(p)=-3ig^2 \fr {\mu^2}{2M_W^2}.$$
On the other hand, a pair of 3-$\phi$-vertices may also make a Higgs-inserter.

4. The ghost-inserters:

For the ghost-inserter in QCD, for example, we take a tree graph with two
ghost-gluon vertices linked by a ghost line with two gluon lines carrying
zero momenta  whose Lorentz indices and color indices are contracted by the
spacetime
metric and the Killing-Cartan metric respectively. Its Feynman rule is given by
$$I^{\{gh\}}_{~~~a_1a_2}(p) = -ig_c^2 C_2({\bf 8}) \delta_{a_1a_2}.$$ For other
ghost-inserters, it is easy to construct in a similar way.

\vskip 7mm
{\it  The work by HYG was mostly done during his visiting to The
Max-Planck-Institut
f\"ur Mathematik, Bonn. He would like to thank Professors F. Hirzebruch  for
warm hospitality. He is also grateful to Professor
W. Nahm for valuable discussion and warm hospitality.
 HYG is supported
in part
by The National Natural Science Foundation of China.}

\vskip 8mm

\bigskip\bigskip
{\raggedright \bf References\\}
\begin{description}

\item[1] Zhong-Hua Wang  and Han-Ying Guo, Intrinsic vertex regularization and
renormalization in $\phi^4$ theory. 1992. ITP-CAS and SISSA preprint.
 Unpublished.

\item[2] Zhong-Hua Wang  and Han-Ying Guo, Intrinsic loop regularization and
renormalization in $\phi^4$ theory. Comm. Theor. Phys. ( Beijing ) {\bf 21}
(1994) 361.

\item[3] Zhong-Hua Wang  and Han-Ying Guo, Intrinsic loop regularization and
renormalization in QED. To appear in Comm. Theor. Phys.  ( Beijing ).

\item[4] Zhong-Hua Wang  and Luc Vinet, Triangle anomaly from the point of view
of loop regularization. 1992. Univ. de Montr\`eal preprint. Unpublished.

\item[5] Dao-Neng Gao, Mu-Lin Yan and Han-Ying Guo,  Intrinsic loop
regularization in quantum field theory. To appear in the Proc. of ITP Workshop
 on QFT (1994).

\end{description}

\end{document}